\documentclass[reprint,footinbib,amsmath,amssymb,aps,prl,floatfix,twocolumn,nobalancelastpage, superscriptaddress]{revtex4-2}

\usepackage{physics}
\usepackage{printlen} % to determine the dimensions of your document
\usepackage{amsthm}
\usepackage{mathtools}
\usepackage{bbm}

\usepackage{xfrac}

\usepackage{graphicx}% Include figure files
\graphicspath{{images/}}
\usepackage{wrapfig}

\usepackage{dcolumn}% Align table columns on decimal point
\usepackage{bm}% bold math
\usepackage[hidelinks]{hyperref}% add hypertext capabilities
\usepackage{tikz}
\usepackage{dsfont}
\usetikzlibrary{quantikz}

\usepackage{diagbox}

\usepackage{lipsum} 

\usepackage[labelformat=simple]{subcaption}

\usepackage{ragged2e}
\DeclareCaptionJustification{justified}{\justifying}
\usepackage{printlen}

\usepackage{etoolbox}

\usepackage[innercaption]{sidecap}
\sidecaptionvpos{figure}{c}

\usepackage{siunitx}
\AtBeginDocument{\RenewCommandCopy\qty\SI}
\makeatletter
\renewcommand{\p@subsubsection}{}
\makeatother

\makeatletter
\def\maketitle{
\@author@finish
\title@column\titleblock@produce
\suppressfloats[t]}
\makeatother

\newcounter{SMsections}

\newcommand{\cl}[1]{\mathcal{#1}}
\newcommand{\be}{\begin{equation}} 
\newcommand{\ee}{\end{equation}}

\newcommand{\ped}[1]{_{\mathrm{#1}}}
\newcommand{\de}{\textrm{d}}
\NewDocumentCommand{\vop}{s m}{%
  \IfBooleanTF{#1}{\widehat{\mathbf{#2}}}{\hat{\mathbf{#2}}}%
}

\begin{document}
\preprint{APS/123-QED}

%%%%%%%%%%%%%%%%%%%%%%%%%%%%%%%%%%%%%%%%%%%%%%%%%%%%%
%%%%%------------------ Title ------------------%%%%%
%%%%%%%%%%%%%%%%%%%%%%%%%%%%%%%%%%%%%%%%%%%%%%%%%%%%%

\title{Modulation theory formulation of atomic light-matter interaction}

%%%%%%%%%%%%%%%%%%%%%%%%%%%%%%%%%%%%%%%%%%%%%%%%%%%%%
%%%%%----------------- Authors -----------------%%%%%
%%%%%%%%%%%%%%%%%%%%%%%%%%%%%%%%%%%%%%%%%%%%%%%%%%%%%

\author{Matteo Simoni}
    \email{masimoni@phys.ethz.ch}
    %\thanks{These authors contributed equally.}
    \affiliation{Institute for Quantum Electronics, ETH Z\"urich, Otto-Stern-Weg 1, 8093 Z\"urich, Switzerland}
    \affiliation{Quantum Center, ETH Z{\"u}rich, 8093 Z{\"u}rich, Switzerland}
\author{Ivan Rojkov}
    \email{ivan.rojkov@yale.edu}
    %\thanks{These authors contributed equally.}
    \affiliation{Yale Quantum Institute, Yale University, New Haven, Connecticut 06520, USA}
    \affiliation{Department of Physics, Yale University, New Haven, Connecticut 06520, USA}
\author{Jonathan Home}
    \email{jhome@phys.ethz.ch}
    \affiliation{Institute for Quantum Electronics, ETH Z\"urich, Otto-Stern-Weg 1, 8093 Z\"urich, Switzerland}
    \affiliation{Quantum Center, ETH Z{\"u}rich, 8093 Z{\"u}rich, Switzerland}

\date{\today}

%%%%%%%%%%%%%%%%%%%%%%%%%%%%%%%%%%%%%%%%%%%%%%%%%%%%%
%%%%%----------------- Abstract ----------------%%%%%
%%%%%%%%%%%%%%%%%%%%%%%%%%%%%%%%%%%%%%%%%%%%%%%%%%%%%
 
\begin{abstract}
\medskip \noindent
We provide a re-formulation of the light-matter interaction of trapped-atom systems in terms of classical modulation theory. We introduce commuting ``mean'' quadrature operators together with ``deviation'' operators that describe the quantum fluctuations resulting from the uncertainty principle. From the  ``mean'' position operator stems an accurate approximate expression for the internal transition coupling strengths in terms of Bessel functions which matches that of classical modulation theory. The error of the approximation is a direct result of quantum fluctuations. We also show that this result can also be obtained with WKB theory. The validity of our approach is numerically verified and supported by an expansion of the exact expression using a recurrence relation between orthogonal polynomials. 
Compared to the exact solution, our result is analytically more tractable, numerically more stable, and admits a transparent physical interpretation which connects the classical and quantum pictures.
\end{abstract}

\maketitle

\noindent 
Trapped atoms such as ions or tweezer arrays are among the most well-studied quantum mechanical systems, in large part thanks to their excellent isolation and pristine control through the use of laser fields \cite{bruzewicz_progress_2019, kaufman_quantum_2021}. Atoms are trapped in optical and electro-magnetic traps, providing external center of mass degrees of freedom which for small excursions can be considered as a harmonic oscillator, alongside their internal states. These systems have been extensively used for quantum computing \cite{cirac_quantum_1995, kaufman_entangling_2015, de_neeve_error_2022, matsos_universal_2025, shaw_erasure_2025}, quantum state engineering \cite{wineland_generation_1996, monroe_schrodinger_1996, kaufman_cooling_2012, fluhmann_encoding_2019, behrle_phonon_2023, brown_time-of-flight_2023, simoni_NLRE_2025, lienhard_generation_2025}, quantum simulation \cite{porras_effective_2004, hartke_register_2022, katz_programmable_2023, magoni_molecular_2023, navickas_experimental_2025, sun_simulation_2025} and precision measurement \cite{schmidt_spectroscopy_2005, mccormick_quantumenhanced_2019, wolf_motional_2019, gilmore_quantumenhanced_2021, valahu_enhanced_2025}. Laser control can be understood through the light-matter interaction, which, under well-justified approximations, can be considered to near-resonantly couple two internal levels $\ket{g}$ and $\ket{e}$ and a single motional oscillation mode of frequency $\omega\ped{o}$. For a plane-wave light field with frequency $\omega\ped{L}$, this Hamiltonian can be written as~\cite{leibfried_dynamics_2003}
\begin{equation} \label{eq:light_matter_ham}
    \hat{\mathcal{H}}=\frac{\hbar\Omega_0}{2}\hat\sigma_+\,
    \exp\!\Big\{i\eta \big(\hat{a}^{\dagger} e^{i\omega\ped{o} t} + \hat{a} e^{-i\omega\ped{o} t}\big)\Big\}  
    e^{-i \delta t} %-\phi)} 
    + \mathrm{H.c.},
\end{equation}
where $\Omega_0$ is the bare Rabi frequency, $\delta=\omega\ped{L} - \omega\ped{eg}$ the detuning and $\eta=kz_0$ the dimensionless Lamb-Dicke parameter, which relates the field's wave vector $k$ and the particle's ground state wave packet size $z_0$. A common interpretation of this interaction is that in the rest-frame of the trapped atom the laser field is phase modulated by the oscillation of the center of mass of the atom. By setting the laser detuning $\delta = s\,\omega\ped{o}$, we select the resonant process corresponding to a sideband of order $s$, which results in an oscillation of the population between states $\ket{g,n}$ and $\ket{e,n+s}$. The coupling strength $\Omega_{n+s,n}=\langle n+s|\, \exp\!\{i\eta(\hat{a}+\hat{a}^\dagger)\}\, | n \rangle$ of this process as a function of the oscillator occupation number $n$ is given exactly by~\cite{cahill_ordered_1969, wineland_cooling_1979}
\begin{equation} \label{eq:rabi_laguerre}
    \Omega_{n+s,n}= e^{-\eta^2/2} \, (i\eta)^{s} \, \sqrt{\frac{n!}{(n+s)!}} \, L_{n}^{(s)}(\eta^2),
\end{equation}
where $L_{n}^{(s)}(\eta^2)$ is the generalized Laguerre polynomial. While this equation is exact, it has several drawbacks. First, it is inconvenient for analytical derivations, due to the fact that the Fock number $n$ enters as the order of the polynomial and not the argument. Second, its numerical evaluation is prone to overflow and underflow issues due to the presence of factorials which are numerically unstable~\cite{provaznik_numerical_2022}. Third, it does not provide an intuitive relation to the physics of the modulation of the laser field, despite arising from such a consideration~\cite{wineland_cooling_1979}. This prevents an appreciation of the connection between classical and quantum physics. 

In this Letter, we approach the problem from a new perspective, which naturally produces an approximate yet accurate result:
\begin{equation} \label{eq:rabi_bessel}
    \Omega_{n+s,n} = i^{s}\,J_{s}\!\left(2\eta\sqrt{n+\frac{s+1}{2}}\right) + \mathcal{O}\!\left(n^{-3/4}\right),
\end{equation}
where $J_s(x)$ is the Bessel function of order $s$ with argument $x$. This result addresses the three points raised for Eq.~\eqref{eq:rabi_laguerre}. Analytically, the Fock number $n$ now appears as the argument of a well-studied special function~\cite{watson_treatise_1995}, making the expression more amenable to asymptotic analysis and closed-form manipulations. Numerically, this representation is free from factorials and fast, stable algorithms for the evaluation of Bessel functions are readily available~\cite{bremer_algorithm_2019}. Most importantly, Eq.~\eqref{eq:rabi_bessel} carries rich physical meaning. The first term is a Bessel function, as would be expected from classical phase modulation theory~\cite{lathi_modern_1998}, and the modulation index is found by minimizing the error of the approximation. This approximate result emerges by introducing a new pair of commuting canonical operators. The further error terms quantify the quantum fluctuations which result from non-commuting operators. 
The parameter determining the validity of the approximation is $\eta/\sqrt{\bar n}$, where $\hbar\omega\ped{o}\bar n$ is the mean oscillator energy of the transition. We find that $\eta/\sqrt{\bar n}\ll1$ corresponds to a classical regime where the commuting canonical operator approximation holds and the quantum fluctuations are negligible.

To obtain equation~\eqref{eq:rabi_bessel}, we seek an approximate expression of Eq.~\eqref{eq:rabi_laguerre} that bridges the gap between quantum and classical description of the light-matter interaction. A typical semi-classical approach approximates the creation and annihilation operators as complex numbers plus a small operator-valued fluctuation, $\hat{a}=\alpha + \delta\hat{a}$~\cite{walls_quantum_2025, bachor_guide_2019}. This method however fails because the constant term is diagonal and produces no off-diagonal terms. We instead rewrite the coupling strengths $\Omega_{n+s,n}$ as
\begin{equation}\label{eq:exponential_X0_V}
  \Omega_{n+s,n} = \bra{n+s}\exp\!\big\{i\eta\big(\hat Q + \hat V\big)\big\}\ket{n},
\end{equation}
with the two operators defined as
\begin{align}
  \hat Q &= \sqrt{m}\sum_{k=0}^{\infty}
    \Big(\dyad{k+1}{k}+\dyad{k}{k+1}\Big),  \label{eq:X0}\\
  \hat V &= \sum_{k=0}^{\infty} v(k)
    \Big(\dyad{k+1}{k}+\dyad{k}{k+1}\Big),  \label{eq:V}
\end{align}
where we introduce a positive parameter $m$ and the residual couplings $v(k) = \sqrt{k+1}-\sqrt{m}$. This decomposition allows us to identify two contributions. $\hat Q$ describes hopping processes between neighboring Fock states with constant strength $\sqrt{m}$. It has the same off-diagonal structure as the usual position operator $\hat q$ in the Fock basis, but the square-root dependence of the matrix elements is replaced by a constant $\sqrt{m}$. The operator $\hat V$ captures the residual square-root dependence of the creation and annihilation operators' coupling strengths, which is a consequence of their commutation relation, making Eq.~\eqref{eq:exponential_X0_V} exact. Analogously to $\hat Q$, we can define a new momentum operator with constant matrix elements,
\begin{equation}
   \hat P = i\sqrt{m}\sum_{k=0}^{\infty}
    \Big(\dyad{k+1}{k}-\dyad{k}{k+1}\Big). 
\end{equation}
These operators satisfy $[\hat Q, \hat P] = 2im\dyad{0}{0}$ meaning that, away from Fock space origin, $\hat Q$ and $\hat P$ are classical versions of $\hat q$ and $\hat p$ obeying classical commutation relations.

Using Feynman's formula~\cite{wilcox_exponential_1967}, we  express Eq.~\eqref{eq:exponential_X0_V} as a Dyson series and isolate the contribution of $\hat Q$ as the $0^{\mathrm{th}}$-order term of the series
\begin{equation}\label{eq:dyson_first_order}
\begin{aligned}
     \Omega_{n+s,n} = &\bra{n+s}e^{i\eta\hat Q}\ket{n} +\\
     \quad i\eta\int_0^1\!\!\mathrm{d}t\,
     &\bra{n+s}e^{i\eta(1-t)\hat Q}\hat V e^{i\eta t\hat Q}\ket{n}
    + \mathcal{O}\!\left(\hat V^2\right)\!.
\end{aligned}
\end{equation}
One can show (End Matter) that the $0^{\mathrm{th}}$-order leads to the result
\begin{equation}\label{eq:dyson_0}
    \langle n+s|e^{i\eta\hat Q}|n\rangle=i^sJ_s\big(2\eta\sqrt{m}\big).
\end{equation} 

Nonetheless, since ${v(k)\gg\sqrt{m}}$ for large $k$, it is not obvious that the $0^{\mathrm{th}}$-order term of the series dominates over the higher-order corrections. We prove this by bounding higher-order terms of the series.
Using the resolution of the identity and Eq.~\eqref{eq:dyson_0}, it is straightforward to see that all higher-order terms are proportional to sums over intermediate Fock states of integrals of products of Bessel functions~(End Matter). For ${\abs{r}\gg x}$, the Bessel function $J_{r}(x)$ decays super-exponentially $(x/|r|)^{|r|}$~\cite{*[{See Eqs.~(9.1.31),~(9.2.1) and~(9.3.1) in }] [{}] abramowitz_handbook_1964}, such that we can restrict the sums to the domain $|k-n|<2\eta\sqrt{m}$, i.e. we can ignore paths that stray away from the initial state $n$ further than $2\eta\sqrt{m}$. Within this window, the largest entry of $\hat V$ compares to $\hat Q$ as
\begin{equation}
    \frac{v(n+2\eta\sqrt{m})}{\sqrt{m}} \simeq \frac{2\eta\sqrt{m}}{2m}
    = \frac{\eta}{\sqrt{m}}\,.
\end{equation}
Therefore, in the limit $\eta/\sqrt{m}\ll1$, the operator $\hat V$ can be treated as a small perturbation of $\hat Q$, and the $l$-th term of the Dyson series scales as $\mathcal{O}\big((\eta/\sqrt{m})^l\big)$. To find the ideal expansion point $m$, we impose that  the contribution of the $1^{\mathrm{st}}$-order of the series is minimized. By expanding $v(k)$ around $m$ to first order, we find the ideal expansion point (End Matter)
\begin{equation}\label{eq:nbar}
    m \equiv \bar n = n+\frac{s+1}{2}.
\end{equation}

Combining Eqs.~\eqref{eq:dyson_0} and \eqref{eq:nbar}, we obtain our main result of Eq.~\eqref{eq:rabi_bessel}. We have shown that the approximate expression of $\Omega_{n+s,n}$ as a Bessel function emerges by adopting a new position operator $\hat Q$ that obeys classical commutation relations a part for $n=0$. The strength of this operator is given by the mean energy $\bar n$ of the $\ket{n}$-to-$\ket{n+s}$ transition. The parameter $\eta/\sqrt{\bar{n}}$ is the cornerstone of our perturbative approach since it quantifies the error of the approximation due to the non-commutativity of the conjugate variables. The $\eta/\sqrt{\bar n}\ll1$ limit is facilitated by either $\eta\ll1$, or $\bar n\gg1$. The first case corresponds to a small-kick regime, in which only states with similar $n$ are sampled and there is no sensitivity to the square-root scaling. The second case corresponds to transitions between high-energy states, for which the relative difference between matrix elements exhibiting square-root scaling is smaller. Therefore, the low quantum fluctuations regime is aided by large $\bar n$, contrary to the Lamb-Dicke approximation, which requires $\eta \sqrt{\bar n}\ll 1$.

The expression in Eq.~\eqref{eq:dyson_first_order} provides a way of deriving correction terms to Eq.~\eqref{eq:rabi_bessel} beyond the first order. The evaluation of higher order terms of the Dyson series involves solving more integrals of Bessel functions. These calculations are carried out in the End Matter up to $2^{\mathrm{nd}}$-order, from which we find
\begin{equation}\label{eq:total_rabi_amp_phase}
    \Omega_{n+s,n} = i^s\left(1+\delta A\right)J_s\!\left(\bar \beta\right) + i^sJ'_s\!\left(\bar \beta\right)\delta\beta,
\end{equation}
where we introduce the modulation index ${\bar\beta = 2\eta \sqrt{\bar n}}$ and indicate the derivative of the Bessel function with respect to its argument with $J'_s$. The errors on the amplitude $\delta A$ and phase $\delta\beta$, defined as
\begin{equation}\label{eq:amplitude_phase_errors}
\begin{aligned}
    \delta A&=+\frac{\eta^4}{6\,\bar\beta^2}; \quad\quad
    \delta\beta = -\frac{\eta^4}{6\,\bar\beta} -\frac{(s^2\!-\!1)\,\eta^4}{3\,\bar\beta^3}\,,
\end{aligned}
\end{equation}
describe an underestimation of the overall coupling strength and a modulation index offset, respectively. Equation~\eqref{eq:total_rabi_amp_phase} is an approximation of Eq.~\eqref{eq:rabi_laguerre} that incorporates correction due to quantum fluctuations up to second order in $\eta/\sqrt{\bar{n}}$. For large $\beta$, the asymptotic expression of the Bessel functions reads $J'_s(\beta)\sim \mathcal{O}\!\left(n^{-1/4}\right)$~\cite{abramowitz_handbook_1964}. Combining this with the $\mathcal{O}\!\left(n^{-1/2}\right)$ scaling of the dominant contribution to $\delta\beta$, we obtain that the asymptotic scaling of these corrections amounts to $\mathcal{O}\!\left(n^{-3/4}\right)$. 

\begin{figure}[h!]
    \centering
    \includegraphics[width=\linewidth]{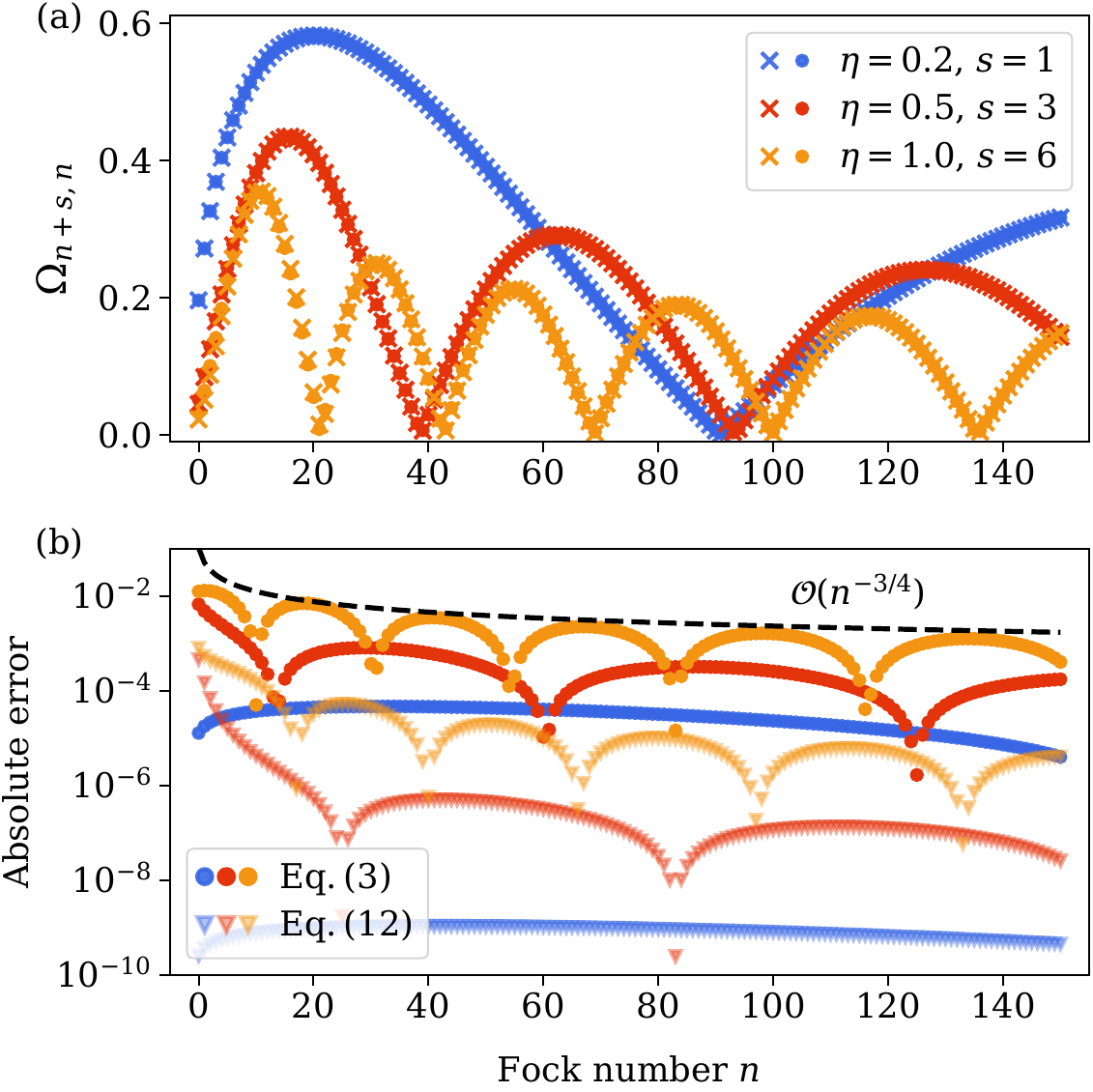}
    \caption{(a) Comparison between the exact matrix elements (crosses, Eq.~\eqref{eq:rabi_laguerre}) and their $0^{\mathrm{th}}$-order approximation (circles, Eq.~\eqref{eq:rabi_bessel}) for $\eta=0.2, 0.5, 1.0$ and $s=1,3,6$. (b) Absolute value of the difference between the exact matrix elements and their approximation. The solid circular markers correspond to the $0^{\mathrm{th}}$-order approximation (Eq.~\eqref{eq:rabi_bessel}), while the fainter triangular markers to the $2^{\mathrm{nd}}$-order (Eq.~\eqref{eq:total_rabi_amp_phase}). The black dashed line shows an envelope following the predicted $\mathcal{O}(n^{-3/4})$ scaling.}
    \label{fig_laguerre_vs_bessel}. 
\end{figure}

We now perform numerical evaluation to put a bound on the error of the approximation. In panel (a) of Fig.~\ref{fig_laguerre_vs_bessel} we present the exact and approximate matrix elements for different $\eta$ and $s$. Panel (b) shows the absolute error between the exact expression and the $0^{\mathrm{th}}$-order (Eq.~\eqref{eq:rabi_bessel}) or the $2^{\mathrm{nd}}$-order (Eq.~\eqref{eq:total_rabi_amp_phase}) approximation formulas. For the $0^{\mathrm{th}}$-order one, the error is consistently below $10^{-2}$. In analogous simulations, we have evaluated the error for a wide range of parameters, $0\le n\le 150$, $0.05\le \eta\le 0.75$ and $0\le s\le 9$, and found a maximum of $2\times10^{-2}$ occurring for $\eta=0.75$, $s=0$, $n=0$~\cite{rojkov_NLRE_2024}. The error of the $2^{\mathrm{nd}}$-order expression is always significantly below $10^{-3}$.
These results cover a broad range of realistic experimental values and motivate the applicability of the approximation n a practical setting. In particular, we have recently adopted it to interpret experimental results of an ion's motional state stabilization even outside of the Lamb-Dicke regime~\cite{simoni_NLRE_2025}.

A classical harmonic oscillation $z(t) = A\cos(\omega\ped{o}t)$ results in the modulation index $\beta = kA$. Comparing this with the expression $\bar\beta = 2\eta \sqrt{\bar n}$ suggests introducing the semi-classical oscillation amplitude $A = 2z_0\sqrt{\bar n}$. We find that this has a clear correspondence to the concept of \textit{turning points} from WKB theory. Turning points are found by imposing that the total energy of the system is equal to its potential energy. Classically, they correspond to spatial positions at which the particle changes oscillation directions. Quantum mechanically, they mark the boundary between the oscillatory behavior of the wave function and quantum exponential decay~\cite{uzer_analytical_1980, child_semiclassical_1991, schleich_phase_2001}. By imposing $\hbar\omega\ped{o} \bar n = \frac{1}{2}m\omega\ped{o}^2 A^2$ we indeed recover the suggested expression for $A$. This means that the approximate expression in Eq.~\eqref{eq:rabi_bessel} can also be obtained with classical modulation theory, using the turning points from WKB theory as the atomic oscillation amplitude. In the End Matter we show that a formal WKB treatment of the light-matter interaction for a trapped atom also yields Eq.~\eqref{eq:rabi_bessel}. By including Airy corrections to the WKB wavefunctions following Ref.~\cite{zambrano_caustic_2008} we are able to recover the dominant correction term in $\delta\beta$ from Eq.~\eqref{eq:amplitude_phase_errors}. 
Our mean-operator approach and the formal WKB treatment can be thought of as complementary: the first one evaluates the matrix element $\Omega_{n+s,n}$ using an approximate operator and exact states, while the latter adopts exact operator but approximate states. The error of the approximation emerges through a quantum-fluctuation operator describing the correct scaling of the ladder operators in the first approach and as the divergences of the WKB wavefunctions at the turning points in the second one.

Our previous derivations rely on physically motivated simplifications of ${\matrixel{n+s}{\exp\{i\eta(\hat{a}+\hat{a}^\dagger)\}}{n}}$. However, the approximation in Eq.~\eqref{eq:rabi_bessel} can also be obtained from Eq.~\eqref{eq:rabi_laguerre} directly. Szeg\"o's Theorem 8.22.4 states that~\cite{szego_orthogonal_1939}
\begin{equation} \label{eq:szegos_thm}
    \frac{x^{s/2}}{e^{x/2}}L_{n}^{s}(x) \simeq
    \frac{(n+s)!}{n!(n+\tfrac{s+1}{2})^{s/2}}\,J_{s}\left(2\sqrt{x}\sqrt{n+\tfrac{s+1}{2}}\right)
\end{equation}
for $x=\eta^2$ and fixed $s$, up to an error $\mathcal{O}\!\left(n^{\,s/2-3/4}\right)$. Combining this expression with the Stirling's approximation of $\sqrt{n!/(n+s)!}$ which is bounded by $\mathcal{O}\!\left(n^{-s/2}\right)$, we obtain the desired approximation in Eq.~\eqref{eq:rabi_bessel} with the same $\mathcal{O}\left(n^{-3/4}\right)$ error scaling. For the detailed mathematical derivation, we refer the reader to Ref.~\cite{*[{}] [{ See Appendix~F for the derivation of the $0^{\mathrm{th}}$-order expression using Szeg\"o's theorem.}] rojkov_NLRE_2024}. To derive the higher-order approximation, we utilize Szeg\"o's recurrence relation used to prove the asymptotic theorem in Eq.~\eqref{eq:szegos_thm}. As before, the calculation involves solving integrals of Bessel functions which we present in the End Matter. Importantly, the obtained expression (see Eq.~\eqref{eq:ivan_error}) matches exactly Eq.~\eqref{eq:total_rabi_amp_phase}. These formal results further motivate the analytical soundness of our previous derivations, both in finding the approximate coupling strengths and the higher-order approximations.

Our approach reformulates the light-matter interaction for trapped atoms from the standpoint of modulation theory. 
We have introduced a position operator that commutes with its momentum counterpart and identified the regime $\eta/\sqrt{\bar n}\ll  1$ as the one in which the light-matter interaction can be satisfactorily approximated by it. To the best of our knowledge, this regime had never been identified before, and justifies a more in-depth study of the high-quantum fluctuations regime. 
We have shown that the adopted commuting operators approach is a powerful way to distinguish classical dynamics from the effect of quantum uncertainty. 
We envision that the approximate but accurate expression derived here will be of utility for the numerical and especially analytical treatment of trapped-atom systems operated in the non-linear regime of the light-matter interaction~\cite{egli_carrier_2026, simoni_NLRE_2025} thanks to the intuitive nature of the modulation theory framework and to the vast range of mathematical results readily applicable to Bessel functions.
The application of the mean-operator approach implemented here to other non-linear functions could offer a potentially powerful method for approximating non-linear Hamiltonians. One example is the Josephson-junction Hamiltonian, which exhibits a cosine dependence on the superconducting phase~\cite{blais_circuit_2021}. The mean-operator approach could enable the analytical investigation of non-linear regimes of operations, past the well-studied low-excitation linear regimes such as the Lamb-Dicke regime of trapped ions and the transmon regime of superconducting circuits. The operation of quantum systems in new regimes of operation across several platforms promises additional rich forms of quantum controls, with consequent developments across quantum technologies.

\medskip\noindent\textbf{Acknowledgments}\\
We would like to acknowledge M. Mazzanti and G. Giuli for stimulating discussions and F.Schmid for feedback on the manuscript.
This work has received funding from the Swiss State Secretariat for Education, Research and Innovation (SERI) – Grant no.: UeM029-6.1. IR acknowledges funding from the Swiss National Science Foundation (Postdoc.Mobility Grant No.~P500-2$\_$235497).

\medskip\noindent\textbf{Competing interests}\\
The authors declare no competing interests.

%%%%%--------------- Bibliography --------------%%%%%

\expandafter\ifx\csname url\endcsname\relax
  \def\url#1{\texttt{#1}}\fi
\expandafter\ifx\csname urlprefix\endcsname\relax\def\urlprefix{URL }\fi
\providecommand{\bibinfo}[2]{#2}
\providecommand{\eprint}[2][]{\url{#2}}

\bibliography{references}

%%%%%--------------- Bibliography Fixed--------------%%%%%

\clearpage
\onecolumngrid
\section{End Matter}
\twocolumngrid

\medskip\noindent
\textbf{$0^{\mathrm{th}}$-order term of the Dyson series.} The mean operator can be conveniently represented in the Susskind–Glogower ladder operators~\cite{susskind_quantum_1964}, $\hat Q=\sqrt{m}\big(\hat Q_++\hat Q_-\big)$, which in the normal order commute everywhere except the origin, $[\hat Q_+,\hat Q_-]=I-\dyad{0}$. Yet, in the limit of small $\eta/\sqrt{m}$, we can consider the sum over $k$ in the definition of $\hat Q$ (see Eq.~\eqref{eq:X0}) to run from $-\infty$ to $+\infty$, and simplify the calculation
\begin{equation*}
\begin{split}
    &\langle n+s|e^{i\eta\hat Q}|n\rangle = 
    %\langle n+s|e^{i\eta\sqrt{m}\,\hat q_+^\dagger}e^{i\eta\sqrt{m}\,\hat q_-}| n\rangle \\
     \sum_{r,l=0}^\infty\frac{\big(i\eta\sqrt{m}\big)^{r+l}}{r!\,l!}\, 
    \langle n+s| (\hat Q_+)^r (\hat Q_-)^l |n\rangle \\ &=\sum_{r=0}^\infty\frac{\big(i\eta\sqrt{m}\big)^{2r+s}}{r!(r+s)!} = i^sJ_s\big(2\eta\sqrt{m}\big)\,.
\end{split}
\end{equation*}

\medskip\noindent
\textbf{Finding the optimal expansion point $m$.}
To find the optimal value of \(m\), we keep the leading term in the
expansion of \(v(k)\) around \(m\),
\begin{equation*}
    v(k)=\frac{k+1-m}{2\sqrt m}
    +\mathcal O\!\left(\frac{(k+1-m)^2}{m^{3/2}}\right).
\end{equation*}
%from which
%\begin{equation*}
%    \hat V \simeq \sum_{k=0}^{\infty}\frac{k+1-m}{2\sqrt m}\Big(
%        \ket{k+1}\!\bra{k} + \ket{k}\!\bra{k+1} \Big).
%\end{equation*}
Inserting this expression into the first-order Dyson term we obtain
\begin{align*}
    \Omega_{n+s,n}^{1^{\mathrm{st}},\,\mathrm{lin}}
    &= i^s\frac{\eta}{2\sqrt m}\int_0^1\!dt\sum_{r}(n+r+1-m)\\
    &\Big[J_{s-r-1}\left(2\eta\sqrt{ m}(1-t)\right)J_r\left(2\eta\sqrt{m} t\right)\\
    & -J_{s-r}\left(2\eta\sqrt{m}(1-t)\right)J_{r+1}\left(2\eta\sqrt{m} t\right)\Big].
\end{align*}
Using the Bessel addition theorem, and performing the integral over $t$, we obtain
\begin{equation*}
    \Omega_{n+s,n}^{1^{\mathrm{st}},\,\mathrm{lin}} = i^s\frac{\eta}{\sqrt m}\left(n+\frac{s+1}{2}-m\right)J_s'(2\eta\sqrt m).
\end{equation*}
Therefore the leading contribution in \(\hat V\) is canceled by choosing $m= n+\frac{s+1}{2}$.

\medskip\noindent
\textbf{$1^{\mathrm{st}}$-order term of the Dyson series.}
To find its explicit contribution, we must solve the integral in Eq.~\eqref{eq:dyson_first_order}.
%\begin{equation} \label{eq:dyson_first_order_sol}
%\begin{aligned}
%    &\Omega_{n+s,n}^{1^{\mathrm{st}}} = i\,\eta\int_0^1\!\mathrm{d}t\;
%    \bra{n+s}e^{i\eta(1-t)\hat Q}\;\hat V\;e^{i\eta t\hat Q}\ket{n} \\
%    &= i^s\eta \int_0^1\!\mathrm{d}t \sum_{j=-n}^{+\infty} v(n+j) \Big[J_{s-j-1}\big((1-t)\bar \beta\big)J_{j}\big(t\bar \beta\big)\\[-5pt]
%    &\qquad\qquad\qquad\qquad\qquad\,\,- J_{s-j}\big((1-t)\bar \beta\big)J_{j+1}\big(t\bar \beta\big)\Big]
%\end{aligned}
%\end{equation}
We do this by using the $0^\text{th}$-order result and expanding $v(n+j)$ in powers of $\bar n^{-1/2}$,
\begin{equation*} %\label{eq:vj_expansion}
\begin{split}
    &v(n+j) %= \sqrt{n+j+1}-\sqrt{\bar n} \\
    =\frac{2j + 1-s}{4\,\bar n^{1/2}} - \frac{(2j +1-s)^2}{32\,\bar n ^{3/2}} + \mathcal{O}\big(\bar n^{-5/2}\big).
\end{split}
\end{equation*}
As we have proven before, the linear term contributes trivially to $\Omega_{n+s,n}^{1^{\mathrm{st}}}$. The quadratic term gives us the leading-order contribution to $\Omega_{n+s,n}^{1^{\mathrm{st}}}$. %Explicitly,
%\begin{equation*}
%\begin{aligned}
%&\Omega_{n+s,n}^{1^{\mathrm{st}}}
%=-\frac{i^s\eta}{32\,\bar n^{3/2}}\int_0^1\!\mathrm{d}t\sum_{j=-n}^{+\infty}(2j+1-s)^2\times\\
%&\Big[J_{s-1-j}\big((1-t)\bar\beta\big)J_j\big(t\bar\beta\big)-J_{s-j}\big((1-t)\bar\beta\big)J_{j+1}\big(t\bar\beta\big)
%\Big].
%\end{aligned}
%\end{equation*}
%Using the same Bessel-product identities as in the zeroth-order calculation and
%\begin{equation*}
%\begin{aligned}
%&\Omega_{n+s,n}^{1^{\mathrm{st}}}=\frac{-i^s\eta}{32\,\bar n^{3/2}}\int_0^1\!\mathrm{d}t\Big\{t^2\bar\beta^2
%\big[J_{s-3}(\bar\beta)+J_{s-1}(\bar\beta)
%-J_{s+1}(\bar\beta)\\&-J_{s+3}(\bar\beta)\big]+2t\bar\beta\big[(2-s)J_{s-2}(\bar\beta)+(2+s)J_{s+2}(\bar\beta)\big]\\
%&-4sJ_{s+1}(\bar\beta)+(1-s)^2\big[J_{s-1}(\bar\beta)-J_{s+1}(\bar\beta)\big]\Big\}.
%\end{aligned}
%\end{equation*}
After performing the integral over $t$, one obtains
\begin{equation*}
\begin{aligned}
&\Omega_{n+s,n}^{1^{\mathrm{st}}}=-\frac{i^s\eta}{32\,\bar n^{3/2}}
\Bigg\{\frac{\bar\beta^2}{3}\big[J_{s-3}+J_{s-1}-J_{s+1}-J_{s+3}\big]\\
&+2\bar\beta\big[J_{s-2}+J_{s+2}\big]-s\bar\beta\big[J_{s-2}-J_{s+2}\big]\\
&-4sJ_{s+1}+(1-s)^2\big[J_{s-1}-J_{s+1}\big]\Bigg\}.
\end{aligned}
\end{equation*}
Finally, reducing the remaining Bessel functions by recurrence relations gives
\begin{equation}\label{eq:first_order_correction}
    \Omega_{n+s,n}^{1^{\mathrm{st}}} = i^s\left[
    \frac{\eta^4}{3\,\bar\beta^2}\,J_s(\bar\beta)
    - \frac{(s^2-1)\,\eta^4}{6\,\bar\beta^3}\,J'_s(\bar\beta)
  \right]\,.
\end{equation}
The $1^{\mathrm{st}}$-order of the Dyson series therefore results in an error proportional to $J_s(\bar\beta)$ and in one proportional to the derivative $J'_s(\bar\beta)$. %Given that we derived these terms from the quadratic expansion of $v(n+j)$ in the $1^{\mathrm{st}}$-order of the Dyson series, we also need to consider the linear expansion of $v(n+j)$ in the $2^{\mathrm{nd}}$-order of the Dyson series.

\medskip\noindent
\textbf{$2^{\mathrm{nd}}$-order term of the Dyson series.}
The $2^{\mathrm{nd}}$-order term of the Dyson series is
\begin{equation}\label{eq:dyson_second_order}
   \begin{aligned} 
    &\Omega^{2^{\mathrm{nd}}}_{n+s,n}=(i\eta)^2 \int_0^1\!\mathrm{d}t_1 \int_0^{t_1}\!\mathrm{d}t_2\\
    &\qquad\,\,\,\,\bra{n+s}
    e^{i\eta(1-t_1)\hat Q}\hat V
    e^{i\eta(t_1-t_2)\hat Q}\hat V
    e^{i\eta t_2\hat Q}
    \ket{n}\!.
    \end{aligned}
\end{equation}
We again expand $v(n+j)$ in powers of $\bar n^{-1/2}$, this time only keeping the linear term since we are at $2^{\mathrm{nd}}$-order in the Dyson series. After expanding the two insertions of $\hat V$ in the Fock basis and using the same Bessel-product identities as before, we can perform the integrals over $t_1$ and $t_2$, obtaining
%the remaining integral can be written as
%\begin{equation*}
%\begin{aligned}
%&\Omega^{2^{\mathrm{nd}}}_{n+s,n}
%=-\frac{i^s\eta^4}{\bar\beta^2}\int_0^1\!\mathrm{d}t_1\int_0^{t_1}\!\mathrm{d}t_2\Bigg\{\\
%&-\frac{(s-1)(s-3)}{4}J_{s-2}+\frac{s^2+1}{2}J_s-\frac{(s+1)(s+3)}{4}J_{s+2}\\
%&+\frac{\bar\beta t_1}{2}\left[\frac{s-1}{2}\bigl(J_{s-3}-J_{s+1}\bigr)-\frac{s+1}{2}\bigl(J_{s-1}-J_{s+3}\bigr)\right]\\
%&+\frac{\bar\beta t_2}{2}\left[\frac{s-5}{2}J_{s-3}+\frac{3-s}{2}J_{s-1}-\frac{s+3}{2}J_{s+1}+\frac{s+5}{2}J_{s+3}\right]\\
%&+\frac{\bar\beta^2 t_1t_2}{4}\left[-J_{s-4}+2J_s-J_{s+4}\right]\Bigg\},
%\end{aligned}
%\end{equation*}
%where all Bessel functions are evaluated at $\bar\beta$. Performing the integrals over $t_1$ and $t_2$ gives
\begin{equation*}
\begin{aligned}
&\Omega^{2^{\mathrm{nd}}}_{n+s,n}=-\frac{i^s\eta^4}{\bar\beta^2}\Bigg\{\\
&-\frac{(s-1)(s-3)}{8}J_{s-2}+\frac{s^2+1}{4}J_s-\frac{(s+1)(s+3)}{8}J_{s+2}\\
&+\frac{\bar\beta}{6}\left[\frac{s-1}{2}\bigl(J_{s-3}-J_{s+1}\bigr)-\frac{s+1}{2}\bigl(J_{s-1}-J_{s+3}\bigr)\right]\\
&+\frac{\bar\beta}{12}\left[\frac{s-5}{2}J_{s-3}+\frac{3-s}{2}J_{s-1}-\frac{s+3}{2}J_{s+1}+\frac{s+5}{2}J_{s+3}\right]\\
&+\frac{\bar\beta^2}{32}\left[-J_{s-4}+2J_s-J_{s+4}\right]\Bigg\}.
\end{aligned}
\end{equation*}
Finally, reducing the shifted Bessel functions to $J_s(\bar\beta)$ and $J'_s(\bar\beta)$ by recurrence relations, all higher-order shifts cancel and one obtains
\begin{equation*}
    \Omega_{n+s,n}^{2^{\mathrm{nd}}} = -i^s\left[
    \frac{\eta^4}{6\,\bar\beta^2}\,J_s(\bar\beta)
    + \frac{(\bar\beta^2+s^2-1)\,\eta^4}{6\,\bar\beta^3}\,J'_s(\bar\beta)
  \right].
\end{equation*}
Combining the $0^{\mathrm{th}}$, $1^{\mathrm{st}}$ and $2^{\mathrm{nd}}$-order contributions, we obtain Eq.~\eqref{eq:total_rabi_amp_phase}.
%quadratic contribution to the $1^{\mathrm{st}}$-order of the Dyson series (Eq.~\eqref{eq:first_order_correction}) with the linear contribution to the $2^{\mathrm{nd}}$-order of the Dyson series (Eq.~\eqref{eq:second_order_correction}) we obtain our final result for the absolute error between Eq.~\eqref{eq:rabi_laguerre} and Eq.~\eqref{eq:rabi_bessel}:
%\begin{equation}\label{eq:total_rabi_from_dyson_app}
%\begin{aligned}
%    &\text{AbsErr}=\abs{\Omega_{n+s,n} - i^sJ_s\big(\bar\beta\big)} 
%    = \abs{\Omega_{n+s,n}^{1^{\mathrm{st}}} + \Omega_{n+s,n}^{2^{\mathrm{nd}}}} \\
%    &=
%    \frac{\eta^4}{6\,\bar\beta^2}\,J_s(\bar\beta)
%    -\frac{\eta^4}{6\,\bar\beta}\,J'_s(\bar\beta)
%    - \frac{\eta^4\,(s^2\!-\!1)}{3\,\bar\beta^3}\,J'_s(\bar\beta).
%\end{aligned}
%\end{equation}

\medskip\noindent
\textbf{WKB treatment light-matter interaction.} 
Here, we show that Eq.~\eqref{eq:rabi_bessel} can be derived with a complete WKB treatment of the light-matter interaction. We want to calculate
\begin{equation}
\label{eq:integral}
    \bra{n+s} e^{i\eta(\hat{a}+\hat{a}^\dagger)} \ket{n} = 
    \int\!\mathrm{d}q\,\, \psi_{n+s}^*(q)\,e^{ikq}\,\psi_n(q).
\end{equation}
Within the WKB method, the wavefunction of the quantum harmonic oscillator with eigenvalue $n$ reads
\begin{equation}
    \label{eq:WKB_wavefunction}
    \psi_n(q) = \sqrt{\frac{1}{\pi}\frac{2m\omega\ped{0}}{p_n(q)}}\,\cos\phi_{n}(q),
\end{equation}
where ${p_n(q)=m\omega\ped{0}\sqrt{q_n^2-q^2}}$ is the quantized momentum and $\phi_{n}(q)$ the phase function
\begin{equation*}
    \phi_n(q) = \left(n+\frac{1}{2}\right)\left(\arccos\frac{q}{q_n}-\frac{q}{q_n}\sqrt{1-\frac{q^2}{q_n^2}}\right) - \frac{\pi}{4}, 
    %&= \frac{E_n}{\hbar\omega\ped{0}}\left(\arccos{x\sqrt{\frac{m\omega\ped{0}^2}{E_n}}- x\sqrt{\frac{m\omega\ped{0}^2}{E_n}}\sqrt{1-\frac{x^2m\omega\ped{0}}{E_n}}}\right) - \frac{\pi}{4}.
\end{equation*}
both of which depend of the so-called turning points, ${q_n = \sqrt{2E_n/(m \omega\ped{0}^2)}}$. The phase function is the classical action normalized to $\hbar$, i.e. $\phi_n(q)=\tfrac{1}{\hbar}\int p_n(q)\de q - \tfrac{\pi}{4}$. The wavefunction in Eq.~\eqref{eq:WKB_wavefunction} %corresponds to the simplest Ansatz satisfying simultaneously the Schrödinger's equation and the correspondence principle. Yet, this solution
is valid only far away from the turning points, given that $p_n(q)\rightarrow0$ for $q\rightarrow q_n$ which makes the wavefunctions diverge. The divergence at these pathological points can be resolved using uniform or Airy corrections~\cite{ghatak_airy_1991}. Disregarding such corrections, the spatial integral in Eq.~\eqref{eq:integral} becomes
%\begin{equation*}
%    \overset{\mathrm{WKB}}{\simeq}\frac{2}{\pi}\int\!\mathrm{d}x \frac{e^{ikx} \cos\phi_{n+s}(x)\,\cos\phi_{n}(x)}{\left[\left(x_{n+s}^2-x^2\right)\left(x_n^2-x^2\right)\right]^{1/4}}.
%\end{equation*}
\begin{equation*}
    \overset{\mathrm{RWA}}{\simeq}\frac{1}{\pi}\int \frac{\mathrm{d}x\,e^{ikq}  \cos\left(\phi_{n+s}(q)-\phi_{n}(q)\right)}{\left[\left(q_{n+s}^2-q^2\right)\left(q_n^2-q^2\right)\right]^{1/4}},
\end{equation*}
where we have ignored the $\cos\left(\phi_{n+s}(q)+\phi_{n}(q)\right)$ integral, which oscillates much faster than the phase difference one and thus averages to zero. 
The phase difference can be approximated as
\begin{equation*}
    \phi_{n+s}(q)-\phi_{n}(q)\simeq \frac{\partial \phi}{\partial E}\bigg\rvert_{\bar{E}}\left(E_{n+s}-E_n\right) = s\arccos{\frac{q}{\bar{q}}},
\end{equation*}
where $\bar{q}=\sqrt{2\bar{E}/(m\omega\ped{0}^2)}$ and $\bar{E}=\tfrac{E_n+E_{n+s}}{2}$. Then, to first order, we obtain
\begin{equation*}
    \overset{\mathrm{LIN}}{\simeq}\frac{1}{\pi}\int\!\mathrm{d}q 
    \frac{e^{ikq}}{\left[\left(q_{n+s}^2-q^2\right)\left(q_n^2-q^2\right)\right]^{1/4}} 
    \cos\!\left(s\arccos{\frac{q}{\bar{q}}}\right).
\end{equation*}
Lastly, we can make our mean-energy approximation, $q_{n+s}\simeq q_n\simeq \bar{q}$, and a change of variable $q=\bar{q}\cos\theta$,
\begin{equation*}
\begin{aligned}
      &\overset{\bar{E}}{\simeq}%\frac{1}{\pi}\int_{-\bar{x}}^{\bar{x}}\!\mathrm{d}x\, \frac{e^{ikx}}{\sqrt{\bar{x}^2-x^2}} \cos\!\left(s\arccos{\frac{x}{\bar{x}}}\right)  \\
      \frac{1}{\pi}\int_{\pi}^{0}\!\mathrm{d}\theta\,(-1)\,e^{ik\bar{q}\cos\theta}\cos\!\left(s\theta\right) = i^s  J_s\!(\bar \beta),
      % &= \frac{1}{\pi}\int_{\pi}^{0}\mathrm{d}\theta\,\frac{-\bar{x}\sin\theta}{\bar{x}\sin\theta}e^{ik\bar{x}\cos\theta}\cos\left(s\theta\right) = i^s  J_s\left(k\bar{x}\right),
\end{aligned}
\end{equation*}
which concludes our derivation. This shows that the Bessel function approximation of $\Omega_{n+s,n}$, obtained in the main text by analogy with classical modulation theory, can also be formally derived from WKB theory.

\medskip\noindent
\textbf{Higher-order terms via quantum-to-classical approach.} 
The derivation of Eq.~\eqref{eq:rabi_bessel} from Eq.~\eqref{eq:rabi_laguerre} involves two approximations~\cite{rojkov_NLRE_2024}: (1) Szeg\"o's Theorem 8.22.4~\cite{szego_orthogonal_1939} and (2) Stirling's formula. To derive the leading error term, we take these approximations to the next order. The theorem is an asymptotic approximation of the Szeg\"o's recurrence relation (see Eq.~(8.64.3) in Ref.~\cite{szego_orthogonal_1939}),
\[
\begin{split}
    e^{-\eta/2} \,\, \eta^{s/2} \,\,L^{(s)}_{n}\big(\eta^2\big) &= \bar n^{-s/2} \,\frac{(n+s)!}{n!}
    \,J_{s}\!\big(\,\bar\beta\,\big) \\
    &-\frac{\pi}{2} \int_0^\eta \!\!\mathrm{d}x\, I(x) \, e^{-x^2/2}\,\,x^{s+3}\,\,L_n^{(s)}\big(x^2\big), \\[-5pt]
\end{split}
\]
where $I(x) = J_s\big(\,\bar{\beta}\,\big)\,Y_{s}\!\left(2x\sqrt{\bar{n}}\right)\! - Y_{s}\big(\,\bar{\beta}\,\big)\,J_s\!\left(2x\sqrt{\bar{n}}\right)$,
with $Y_{s}$ being Bessel functions of second kind. To find the $2^\text{nd}$-order approximation, we substitute the last terms in the integral by the recurrence relation and solve the resulting integral,
\be \label{eq:bessel_integral}
\begin{aligned}
     \frac{\pi}{2}\int_0^{\eta} \!\!\mathrm{d}x&\,I(x)\,x^3\,J_s\!\left(2x\sqrt{\bar{n}}\right) = -\frac{\eta^4}{12\bar\beta}\Big[J_{s - 1}(\bar{\beta}\big) - J_{s + 1}(\bar{\beta}\big)\Big] \\
    &+ \frac{1}{6} \frac{\eta^4}{\bar{\beta}^2}\,J_{s}(\,\bar{\beta}\,\big)
    + \frac{1}{6} \frac{\eta^4}{\bar{\beta}^3}2(s^2-1)\,J_{s+1}(\,\bar{\beta}\,\big),   
\end{aligned}
\ee 
where we identify the first term as the derivative of the Bessel function. Higher-order approximations can be obtained similarly and involve solving nested integrals. A closed form solution may exist but we leave this problem for further works.

The Stirling's formula is used to approximate 
\be \label{eq:sqrt_factorial_approximation}
    \left(\frac{n!}{(n+s)!}\right)^{\!1/2} = 
    \bar{n}^{s/2} \, \frac{n!}{(n+s)!}\,\, E_s(\bar{n}),
\ee
where the error term $E_s(\bar{n})$ reads
\be \label{eq:sqrt_factorial_err}
\begin{split}
    E_s(\bar{n}) &= \left(1-\frac{s(s^2-1)}{24\,\bar{n}^2} + \cl{O}(\bar{n}^{-4})\!\right)^{\!1/2} \\ 
    &\approx 1 - \frac{s(s^2-1)}{48\,\bar{n}^2} = 1 - \frac{1}{6} \frac{\eta^4}{\bar{\beta}^4} 2s (s^2-1).
\end{split}
\ee

Combining Eqs.~\eqref{eq:bessel_integral} and~\eqref{eq:sqrt_factorial_err} with the quantum formula for $\Omega_{n+s,n}$, we get that the absolute error between Eqs.~\eqref{eq:rabi_laguerre} and~\eqref{eq:rabi_bessel} has the following form (up to $\eta^4/\bar{\beta}^4$ order)
\[
\begin{split}
    \text{AbsErr}=&
    -\frac{1}{6} \frac{\eta^4}{\bar{\beta}^4} 2s (s^2-1) \, J_s\big(\,\bar{\beta}\,\big)
    - \frac{1}{6} \frac{\eta^4}{\bar\beta}\,J'_s(\bar\beta) \\
    &+ \frac{1}{6} \frac{\eta^4}{\bar{\beta}^2}\,J_{s}(\,\bar{\beta}\,\big)
    + \frac{1}{6} \frac{\eta^4}{\bar{\beta}^3}2(s^2-1)\,J_{s+1}(\,\bar{\beta}\,\big)\,.
\end{split}
\]
Numerically, we find that each term in this expression is important. Using Bessel function properties, it can be further simplified to
\be \label{eq:ivan_error}
\begin{split}
    \text{AbsErr}&=
    \frac{\eta^4}{6\,\bar\beta^2}\,J_s(\bar\beta)
    -\frac{\eta^4}{6\,\bar\beta}\,J'_s(\bar\beta)
    - \frac{\eta^4\,(s^2\!-\!1)}{3\,\bar\beta^3}\,J'_s(\bar\beta).
\end{split}
\ee

\clearpage
\onecolumngrid
% \section{Supplementary information}
\title{Supplementary information:\\Modulation theory formulation of atomic light-matter interaction}
\maketitle

\setcounter{equation}{0}
\renewcommand{\theequation}{S\arabic{equation}}
\setcounter{figure}{0}
\renewcommand{\thefigure}{S\arabic{figure}}
\setcounter{page}{1}

\medskip\noindent
\textbf{$0^{\mathrm{th}}$-order term as propagator on a chain of spins.} Alternatively, we can interpret $\hat Q$ as an operator acting on a chain of spins. Consider an infinite 1D chain of spin $1/2$ interacting through the following Hamiltonian:
\begin{equation}\label{eq:1d_uniform_hop}
    \hat{\mathcal{H}} = t \sum_{k=-\infty}^{\infty} \left(\sigma_+^{k+1}\sigma_-^k + \sigma_+^{k}\sigma_-^{k+1}\right).
\end{equation}
This Hamiltonian describes nearest neighbor hopping with uniform tunneling rate $t$ and is identical to $\hat Q$, if we identify $t=\sqrt{\bar n}$ and map the Fock state $\ket{k}$ to the spin state $\ket{\downarrow\,\downarrow\ldots\uparrow\ldots\downarrow}$ with $\uparrow$ at position $k$. We can define the plane-wave states:
\begin{equation}\label{eq:plane_wave_state}
    \ket{a} = \sum_k \frac{1}{\sqrt{2\pi}}e^{i a k}\ket{k}.
\end{equation}
We observe that:
\begin{equation*}
    \hat{\mathcal{H}}\ket{a} = t \sum_k \frac{1}{\sqrt{2\pi}}e^{i a k} \left(\ket{k+1}+\ket{k-1}\right) = 2t\cos(a)\ket{a},
\end{equation*}
i.e. the plane wave is an eigenstate of the uniform tunneling Hamiltonian with eigenvalue $2t\cos(a)$. If we now calculate the propagator of the Hamiltonian in \eqref{eq:1d_uniform_hop} between spin states $\ket{n}$ and $\ket{n+s}$ over time $\tau$ we obtain:
\begin{equation*}\label{eq:1d_uniform_propagator}
\begin{aligned}
    \langle n+s|e^{i\mathcal{H}\tau}&|n\rangle = \int_{-\pi}^{\pi}\de a \,\langle n+s|a\rangle e^{i2t\tau\cos(a)} \langle a|n\rangle \\
    &= \int_{-\pi}^{\pi}\de a \frac{1}{2\pi}\, e^{i a(n+s-n)} e^{i2t\tau\cos(a)} \\
    &= \int_{-\pi}^{\pi}\de a \frac{1}{2\pi}\, e^{i s a}\sum_l i^l J_l(2t\tau)e^{i l a} = i^s J_s(2t\tau).
\end{aligned}
\end{equation*}
Where in the last step we have used the Jacobi-Anger identity. Identifying $t=\sqrt{\bar n}$ and $\tau = \eta$ we recover our main result.

\medskip\noindent
\textbf{Derivation of the error as Airy corrections to WKB method}
We are now going to derive an expression for the fluctuations $\delta\beta$ using WKB theory and relying on \cite{zambrano_caustic_2008}. In this work, the authors calculate the matrix element $\langle n |\, e^{ik x}\, | n \rangle$ semi-classically using WKB approximation and including uniform Airy corrections that prevent pathological behavior at the turning point \cite{ghatak_airy_1991}. We can think of this matrix element as measuring the overlap between the initial state and one displaced along momentum by a momentum kick of amplitude $\xi=\hbar k/\sqrt{m\omega_0}$. Eq 6.7 from \cite{zambrano_caustic_2008} shows that $\langle n |\, e^{ik x}\, | n \rangle$ features oscillations as a function of $n$ with frequency set by the overlap between the phase-space area of the state before and after the kick, in units of $2\hbar$. The area of the initial state is that of a circular orbit in phase space with its action $S$ quantized according to the Bohr-Sommerfeld principle, $S=\hbar\left(n+1/2\right)$. The overlap area $\mathcal{A}(\xi)$ can be found geometrically by calculating the overlap of two circles offset by $\xi$ (the geometric shape of a lens). If we calculate this overlap area starting from a state with energy equal to the mean of the $\ket{n}$ to $\ket{n+s}$ transition, we obtain:
\begin{equation}
    \label{eq:area_phase_space}
    \begin{aligned}
    \frac{\mathcal{A}(\xi)}{2\hbar} &= \frac{\pi \bar{S}}{\hbar} - \frac{|\xi|}{2\hbar}\sqrt{2\bar{S}-\left(\frac{|\xi|}{2}\right)^2}-\frac{2\bar{S}}{\hbar}\arcsin\left(\frac{|\xi|}{2\sqrt{2\bar{S}}}\right) \\
    &= \pi \bar{n} - 2\bar{n}\left(\frac{\eta}{2\sqrt{\bar{n}}}\sqrt{1-\frac{\eta^2}{4\bar{n}}} + \arcsin\left(\frac{\eta}{2\sqrt{\bar{n}}}\right)\right),
    \end{aligned}
\end{equation}
where we have introduced $\bar{n}=n+\tfrac{s+1}{2}$, and the corresponding mean action $\bar{S}$.
Expanding Eq.~\eqref{eq:area_phase_space} in powers of $\eta/(2\sqrt{\bar{n}})$ we obtain the following expression for the complementary area between the original circle and the lens:
\begin{equation}
    \label{eq:modulation_index_generalized}
    \frac{2\pi\hbar\bar{n}-\mathcal{A}(\xi)}{2\hbar} \simeq 2\eta\sqrt{\bar{n}} - \frac{\eta^3}{12\sqrt{\bar{n}}} + \mathcal{O}\left(\frac{\eta^5}{\bar{n}^{3/2}}\right).
\end{equation}
Substituting $\bar{n}$ for $n$, this expression corresponds to the oscillation frequency of the $\langle n |\, e^{ik x}\, | n \rangle$ element found in \cite{zambrano_caustic_2008}.
We immediately recognize our usual definition of $\bar{\beta}$ in the first term of \eqref{eq:modulation_index_generalized} and we therefore identify \eqref{eq:modulation_index_generalized} as $\bar{\beta}+\delta\beta'$, with
\begin{equation}
    \label{eq:beta_fluctuations_WKB}
    \delta\beta' = -\frac{\eta^3}{12\sqrt{\bar{n}}} = -\frac{\eta^4}{6\bar{\beta}}.
\end{equation}
We have thus found the first term of the phase error $\delta\beta$ from Eq.~\eqref{eq:amplitude_phase_errors} by considering the fact that the overlap between the area of the state before and after the kick changes non-linearly with the kick amplitude, following the equation of a lens. The first order treatment does not consider this effect, which is most relevant for large kicks (large $\eta$) or small initial states (small $\bar{n}$).

\end{document}